\documentclass[9pt,twocolumn,twoside]{pnas-new}

\templatetype{pnasbriefreport} 

\setboolean{displaywatermark}{false}

\def\iso#1#2{\mbox{${}^{#2}{\rm #1}$}}
\def\o1#1{\iso{O}{1#1}}
\def\fe6#1{\iso{Fe}{6#1}}
\def\sm14#1{\iso{Sm}{14#1}}
\def\u23#1{\iso{U}{23#1}}
\def\pu24#1{\iso{Pu}{24#1}}

\def\msol{M_\odot}
\def\gtsim{\stackrel{>}{\sim}}

\title{Supernova Triggers for End-Devonian Extinctions}

\author[a,b,c,1]{Brian D. Fields}
\author[d]{Adrian L. Melott} 
\author[e,f,g]{John Ellis}
\author[a,b]{Adrienne F. Ertel}
\author[h]{Brian J. Fry}
\author[i,j]{Bruce S. Lieberman}
\author[a,b]{Zhenghai Liu}
\author[a,b]{Jesse A. Miller}
\author[k]{Brian C. Thomas}

\affil[a]{Illinois Center for Advanced Studies of the Universe}
\affil[b]{Department of Astronomy, University of Illinois, 1002 W. Green St., Urbana IL 61801, USA}
\affil[c]{Department of Physics, University of Illinois, Urbana IL 61801, USA}
\affil[d]{Department of Physics and Astronomy, University of Kansas, Lawrence KS 66045, USA}
\affil[e]{Department of Physics, Kings College London, Strand, London WC2R 2LS, UK}
\affil[f]{Theoretical Physics Department, CERN, CH-1211 Geneva 23, Switzerland}
\affil[g]{Laboratory of High Energy and Computational Physics, National Institute for Chemical Physics and Biophysics, R\"{a}vala 10, 10143 Tallinn, Estonia}
\affil[h]{Physics Department, United States Air Force Academy, Colorado Springs, CO 80840, USA}
\affil[i]{Department of Ecology, University of Kansas, Lawrence KS 66045, USA}
\affil[j]{Evolutionary Biology and Biodiversity Institute, University of Kansas, Lawrence KS 66045, USA}
\affil[k]{Department of Physics and Astronomy, Washburn University, Topeka KS 66621, USA}

\leadauthor{Fields} 

\authorcontributions{Author contributions: BDF, JE, and ALM designed the research; BDF, BJF, ZL, and JAM made calculations; ALM, AFE, BSL, and BCT analyzed data; BDF, BJF, JE, BSL, ALM, and BCT wrote the paper.}
\authordeclaration{The authors are unaware of any competing interests.}
\correspondingauthor{\textsuperscript{1}To whom correspondence should be addressed. E-mail: bdfields@illinois.edu}


\begin{abstract}
The Late Devonian was a protracted period of low speciation resulting in biodiversity decline,
culminating in extinction events near the Devonian-Carboniferous boundary. Recent evidence indicates that 
the final extinction event may have coincided with a dramatic drop in stratospheric ozone, 
possibly due to a global temperature rise.
Here we study an alternative possible cause for the postulated ozone drop:  a
nearby supernova explosion that could inflict damage 
by accelerating cosmic rays that can deliver ionizing radiation
for up to $\sim 100$ kyr.
We therefore propose that the end-Devonian extinctions were triggered by 
supernova explosions at $\sim 20 \ \rm pc$, 
somewhat beyond the ``kill distance'' that would have precipitated a full mass extinction.  
Such nearby supernovae are likely due to core-collapses of massive stars; these
are concentrated in the thin Galactic disk where the Sun resides. 
Detecting either of the long-lived radioisotopes
confirm a supernova origin, point to the core-collapse explosion of a massive star, 
and probe supernova nucleosythesis. Other possible tests of the supernova hypothesis are discussed.
\end{abstract}

\dates{This manuscript was compiled on \today}
\doi{https://doi.org/10.1073/pnas.2013774117}

\begin{document}

\maketitle
\thispagestyle{firststyle}
\ifthenelse{\boolean{shortarticle}}{\ifthenelse{\boolean{singlecolumn}}{\abscontentformatted}{\abscontent}}{}


The late Devonian biodiversity crisis 
is characterized by a protracted decline in speciation rate occurring over millions of years~\cite{Stigall2020,Fan2020},
punctuated by an extinction pulse (Kellwasser event) followed  $\sim 10$ Myr later by a more moderate extinction (Hangenberg event) around the Devonian-Carboniferous boundary (DCB) $\sim 359$~Myr ago~\cite{Kaiser2015,Bonda2017}.  
Marshall et al.~\cite{Marshall2020} recently suggested that the Hangenberg event was associated with ozone depletion (see also~\cite{Cockell1999}),
in light of evidence such as malformations persisting in palynological assemblages on the order of many thousands of years.
Ref.~\cite{Racki2020} argued that volcanic eruption and 
a large igneous province (LIP) triggered ozone depletion, whereas~\cite{Marshall2020} instead linked 
it 
to an episode of global warming not caused by LIP. 

Previous work has not 
considered astrophysical sources of ionizing radiation, which are known to be possible causes of ozone depletion and concomitant UV-B increase that could trigger elevated extinction levels (see, e.g.,~\cite{Melott2011}),
{as well as direct genetic damage}. 
Here we consider whether astrophysical sources could account for the data in \cite{Marshall2020}, and whether any additional evidence could test for their occurrence.

The precise patterns prevalent during the DCB are complicated by several factors including difficulties in stratigraphic correlation within and between marine and terrestrial settings and the overall paucity of plant remains~\cite{Prestianni2016}. However, a general consensus seems to be emerging that there was first 
a loss of diversity in spores and pollen
followed
after about 300~kyr~\cite{Myrow2014} by a pulse of extinctions of many plants
including proto-trees, armored fish, trilobites, ammonites, 
conodonts, chitinozoans and acritarchs, possibly coeval with the Hangenberg Crisis; this seems to have largely left intact sharks, bony fish and tetrapods with five fingers and toes.
The fact that these species disappeared over multiple beds indicates that the extinction 
extended over at least thousands of years. 

Refs.~\cite{Filipiak2010,Prestianni2016,Marshall2020} also report the discovery of spores from this episode with distinct morphologies including malformed spines and dark pigmented walls,
features consistent with severely deteriorating environmental conditions, and UV-B damage following destruction of the ozone layer~\cite{Filipiak2010}. However, more quantitative data are needed to study their variation
during quiescent times in the fossil record. 

\section*{Heating Mechanism for Ozone Depletion}

Ref.~\cite{Marshall2020} proposes an ozone depletion mechanism involving increased water vapor in the lower stratosphere caused by enhanced convection due to higher surface temperatures.  Water vapor contributes to a catalytic cycle that converts inorganic chlorine (primarily HCl and ClONO$_{2}$) to free radical form (ClO).  The ClO then participates in an ozone-destroying catalytic cycle.  A similar set of cycles involving Br contributes to ozone depletion, but to a lesser extent~\cite{Anderson2012}.  Increased ClO and decreased ozone following convective injection of water into the lower stratosphere has been verified by observation and modeling~\cite{Anderson2012,Anderson2017}.  Ref.~\cite{Marshall2020} argues that a period of exceptional and sustained warming would lead to the loss of the protective ozone layer via this mechanism.

This mechanism is important for lower stratosphere ozone depletion, and may have consequences for ground-level UV-B exposure~\cite{Anderson2012}.  More detailed study is warranted.  Until then, it is unclear whether this change would be sufficient to cause an extinction.  There are several reasons for this.  

First, the vertical extent of this ozone depletion mechanism should be limited to the lower  stratosphere ($\sim 12-18$ km altitude) and does not overlap with the largest concentration of ozone, which occurs around 20-30 km.  So, while depletion may be significant in the lower stratosphere, the bulk of the ozone layer lies above this region and would not be affected.  The total column density 
would be reduced, but not to the extent of a complete loss of the protective ozone layer. 

Secondly, the duration of the effect should be relatively short, 
$\lesssim 1 \ \rm week$~\cite{Anderson2012}, since the injected water vapor is photolyzed and ClO is converted back to HCl and ClONO$_2$.  Thus, unless convective transport of water vapor to the lower stratosphere, e.g., by storms, is continuous (on 
week timescales) the ozone reduction will be episodic, not sustained.  The effect is also seasonal, since strongly convective storms tend to be limited to the spring/summer. While this is likely detrimental to surface life, most organisms have repair mechanisms that can cope with some short-duration UV-B exposure.

Thirdly, the effect is likely to be limited geographically, since strongly convective storms are not uniformly distributed and the enhanced water vapor is likely only to spread over $\sim 100$~km horizontally~\cite{Anderson2012}.

Finally, there is significant uncertainty as to the ozone depletion level needed to induce aberrations in pollen morphology and even more critically, large-scale extinction.  While the anthropogenic ozone ``hole'' over Antarctica has led to increased UV-B exposure, no crash in the ecosystem has resulted.  This may partly be due to the seasonal nature of the change, as would be the case here as well.  Recent work~\cite{Neale2016} has shown that short-term exposure to significant increases in UV-B does not result in large negative impacts on the primary productivity of ocean phytoplankton, and other organisms show a wide range of sensitivity~\cite{Thomas2015,Thomas2018}.  The amount of column depletion over a given location in those cases was $\sim 50\%$.  The depletion caused by the mechanism considered in~\cite{Marshall2020} seems unlikely to be that large.  
Hence, the convective transport of water vapor to the lower stratosphere may not be sufficient to induce a substantial extinction.  It is thus worth considering other mechanisms for global ozone depletion.

\section*{Astrophysical Agents of Ozone Destruction and Biosphere Damage}

Astrophysical mechanisms for biosphere damage include bolide impacts, solar proton events, supernova (SN) explosions, gamma-ray bursts, and neutron star mergers (kilonovae). 
Bolide impacts, gamma-ray bursts and solar proton events are essentially impulsive, and recovery of the ozone layer takes $\lesssim 10 \ \rm yr$~\cite{Thomas2005}, which is likely to avert lasting biosphere destruction. Moreover, these events and kilonovae are unlikely to recur frequently.  Accordingly, we focus on SNe.

Supernovae (SNe) are prompt sources of ionizing photons: extreme UV, X-rays, and gamma rays.  Over longer timescales, the blast collides with surrounding gas, forming a shock that drives particle acceleration.  In this way, SNe produce cosmic rays, i.e., atomic nuclei accelerated to high energies. 
These charged particles are magnetically confined inside the SN remnant, and 
are expected to bathe the Earth for $\sim 100 \ \rm kyr$.  

The cosmic-ray intensity would be high enough to deplete the ozone layer and induce UV-B damage for thousands of years~\cite{Ruderman1974, Ellis1993, Gehrels2003, Melott2017b}. In contrast to the episodic, seasonal, and geographically limited ozone depletion expected from enhanced convection, ozone depletion following a SN is long-lived and global (see, e.g.,~\cite{Gehrels2003, Melott2017b, Thomas2018}) and is therefore much more likely to lead to an extinction event, even given uncertainties around the level of depletion necessary. (We note that, as well as the induced UV-B damage, cosmic rays could also cause radiation damage via muons produced when they impact the atmosphere~\cite{Melott2017a}). 
The SN blast itself is unlikely to wreak significant damage on the biosphere, but may deposit detectable long-lived nuclear isotopes that could provide distinctive signatures, as we discuss later.

There are two main types of SNe:
(1) massive stars ($\gtsim 8 \msol$) that explode as core-collapse SNe (CCSNe), and (2)  
white dwarfs that accrete from binary
companions and explode as Type Ia SNe.
These SN types have similar explosion energies, and both produce ionizing radiation able to damage the biosphere.
However, their different nucleosynthesis outputs lead to different radioisotope signatures.

Near-Earth CCSNe are more likely than Type Ia SNe.  
We estimate the nearby CCSN frequency using a Galactic rate ${\cal R}_{\rm CCSN}=(30 \ \rm yr)^{-1}$ and placing the Sun at a radius $R_\odot=8.7 \ \rm kpc$ in a thin disk of scale radius 2.9~kpc and height 0.1~kpc \cite{Adams2013}.  This gives a CCSN rate
${\cal R}_{\rm SN} = e^{-R_\odot/R_0} \, r^3/3 R_0^2 h_0 \sim 4 \ r_{20}^3 \ \rm Gyr^{-1}$ within $r_{20} = r/20 \ \rm pc$ from Earth.  Hence a CCSN at a distance $\sim 2$ times the ``kill radius'' of 10~pc is a plausible origin of the end-Devonian event(s).  In contrast, the Type Ia SN rate is an order of magnitude smaller, as these events are spread over the 
$\sim 8$ times larger volume of the thick disk. 

Massive stars are usually born in clusters
(OB associations), and are usually in binaries with other massive stars.  Thus, if one CCSN occurred near the DCB, likely there were others.  This could explain the Kellwasser and other enigmatic Devonian events, in addition to the Hangenberg event.

\section*{Possible Radioisotope Signatures of Supernovae}

A CCSN close enough to cause a significant extinction
would also deliver SN debris to the Earth
as dust grains--micron or sub-micron
sized particles created early after the explosion.
Grains in the explosion would decouple from the plasma (gas)
and propagate in the magnetized SN remnant until they
are stopped or destroyed by sputtering during collisions \cite{Fields2019}.

The portion that reaches the Earth would deposit in the
atmosphere live (undecayed) radioactive isotopes.
There is very little pre-existing background
for radioisotopes whose lifetimes are much shorter than the age of the Earth.
Those with lifetimes comparable to the time since the event would provide suitable signatures.
The discoveries of live \fe60 in the deep ocean, the lunar regolith and 
Antarctic snow provide one such signal, 
which is interpreted as due to at least one recent nearby CCSN 
2--3 Myr ago at a distance $\sim 50-100$ pc, which is compatible with
the rate estimate given above \cite{Fields2019}.

Possible relic SN radioisotopes from the end-Devonian period
with an age 360~Ma
include \sm146 (half-life 103~Myr),  \u235 (half-life 704~Myr) and \pu244 (half-life 80.0~Myr). 
The most promising signature may be provided by
\pu244, which has also been discovered in deep-ocean crust and sediment
samples deposited over the last 25~Myr~\cite{Wallner2015}.
Moreover, it is absorbed into bones and retained during life~\cite{Takizawa1982}, 
whereas uranium is absorbed during 
fossilization~\cite{Koul1979} and \sm146 is soluble. There is a significant
\u235 background surviving from before the formation of the Solar System, 
with $(\u235/\u238)_{\oplus} = 0.721 \pm 0.001\%$,
so a significant detection above this background 
requires deposition attaining
$\u235_{\rm SN}/\u238_{\oplus} \gtsim 3 \times 10^{-5}$.
U-Pb dating has been used to date the end-Devonian extinction, 
but with an uncertainty in the \u235/\u238 ratio that is
much larger than this target sensitivity, but
even a few atoms of non-anthropogenic \pu244 in 
end-Devonian fossils would be unambiguous evidence for the {\em r}-process in SNe.

We have estimated the terrestrial deposition
of \sm146, \u235 and \pu244 by a nearby SN. \sm146 is a proton-rich (``{\em p}-process'')
nucleus that might be produced by CCSNe 
or 
Type Ia SNe~\cite{Arnould2003}. 
Models for the {\em p}-process~\cite{Arnould2003} give $\sm146/\sm144 \sim 0.01-2.5$, with the predicted core-collapse abundance typically around 0.2. Assuming a CCSN that produced a solar $\sm144/\o16$ ratio, and ejected $M_{\rm ej}(\o16)=2\msol$, we estimate a total yield of \sm146 in the ejecta of ${\cal N}(\sm146) \sim 1.6 \times 10^{47}$ atoms.
On the other hand, \pu244 and \u235 are neutron-rich nuclei that
are made by the rapid capture of neutrons, the {\em r}-process, whose
astrophysical sites are uncertain.
There is evidence that kilonovae
make at least {\em some} of the lighter {\em r}-process nuclei~\cite{GBM:2017lvd}, 
but it is uncertain whether these events make the heavier nuclei of interest here.
Assuming that CCSNe are the dominant {\em r}-process sites, we estimate yields of ${\cal N}(\u235,\pu244) \sim (3, 1.6) \times 10^{47}$ atoms per explosion. 

The journey of SN-produced radioisotopes
from explosion to seafloor is complex.
Ejecta in dust most readily reaches the Earth \cite{Fry2015}.
The fraction of atoms in dust $f_{\rm dust}$  
should be high for the refractory species of interest.  
Due to their high speeds, SN dust grains will
easily overcome the solar wind and reach Earth
\cite{Fry2016}.
The fallout on Earth
favors deposition at mid-latitudes;
additional dispersion occurs due to ocean currents
\cite{Fry2016}.
The global-average surface density of isotope $i$
with half-life $t_{1/2}$ 
is 
$N_i = f_{\rm dust} {\cal N}_{{\rm ej},i} 2^{-t/t_{1/2}}/16 \pi r^2$  \cite{Fry2015},
with $t$ the time since the explosion.
We thus find global-averaged
end-Devonian surface densities of SN material
\begin{equation}
    N(\sm146,\u235,\pu244) \sim f_{\rm dust} (1,9,0.3) \times 10^{5} {\rm atoms/cm^2} \ r_{20}^{-2} \nonumber
\end{equation}
after including the decay factors for each species. Unfortunately, this estimate implies a ratio of
SN-produced \u235 to the background level in the Earth's crust of ${\cal O}(10^{-10})$, which is
undetectably small. On the other hand, there is no natural background to the prospective \pu244 signal,
which may be detectable in fossiliferous material. Its detectability depends on the temporal resolution of the available geological sample, 
whereas the possible detectability of the prospective \sm146 signal
depends also on the degree of dilution due to its solubility.
Finally, if more than one SN occurred before the DCB, then each of these could deposit radioisotope signals.

\section*{Other Tests for Supernovae}


Some hundreds or thousands of years after the optical and ionizing outburst, the cosmic-ray and dust bombardment of the Earth would begin, with several possible effects.

Cosmic-ray ionization of the atmosphere and
accompanying electron cascades may lead to more frequent lightning, increased nitrate deposition, and wildfires
\cite{Melott2019}. The increased nitrate flux might have
led to CO$_2$ drawdown via its fertilization effect~\cite{Melott2018},
thereby cooling the climate. There is evidence for
cooling during the first stage of the DCB, though this occurred an estimated
300~kyr before the radiation damage attested by the data on pollen and spores~\cite{Marshall2020}.
Any increases in soot and carbon deposits during the end-Devonian
could have been generated by increases in wildfires
\cite{Melott2019}.

Cosmic rays striking the atmosphere produce 
energetic muons that can penetrate
matter to a much larger depth than UV-B radiation. 
The
radiation dose due to muons at the Earth's surface~\cite{Thomas2016}
and in the oceans at depths 
$\lesssim 1 \ \rm km$~\cite{Melott2017a} could exceed for many
years the current total radiation dose at the Earth’s surface from all sources.
Therefore, in addition to comparing the effects of muons
and UV-B radiation at or near the surface, 
they could be considered in end-Devonian extinctions of megafauna living at depth.

Finally, if there was one CCSN at the DCB, there may have been more, which may have been responsible for the Kellwasser and additional events.  
These could show evidence for ozone depletion and the other signatures above.




\acknow{This work was partially supported by grants from the UK STFC and the Estonian Research Council.}

\showacknow{} 

\bibliography{devo-bib}

\end{document}